\documentclass[journal]{vgtc}                
\ifpdf
  \pdfoutput=1\relax                   
  \usepackage{graphicx}                
  \DeclareGraphicsExtensions{.pdf,.png,.jpg,.jpeg} 
\else
  \usepackage{graphicx}                
  \DeclareGraphicsExtensions{.eps}     
\fi%

\graphicspath{{figures/}{pictures/}{images/}{./}} 

\usepackage{microtype}                 
\PassOptionsToPackage{warn}{textcomp}  
\usepackage{textcomp}                  
\usepackage{mathptmx}                  
\usepackage{times}                     
\usepackage{cite}                      
\usepackage{tabu}                      
\usepackage{tabularx}
\usepackage{booktabs}                  

\usepackage{enumitem}
\setlist[itemize]{noitemsep, topsep=0pt}
\usepackage{wrapfig}
\usepackage{caption}
\usepackage{titlecaps}
\usepackage{dblfloatfix}
\usepackage{amsmath}
\usepackage{amssymb}
\usepackage{newtxmath}
\usepackage{fontawesome}
\usepackage{color}
\usepackage{physunits}

\newcommand{\Omnioculars}{\includegraphics[scale=0.085]{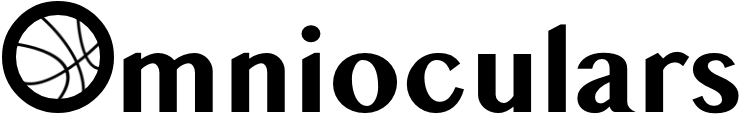}}
\newcommand{\OmniocularsTitle}{\includegraphics[scale=0.14]{images/Omni.png}}%

\DeclareRobustCommand{\inlinefig}[1]{%
\begingroup
\setbox0=\hbox{\includegraphics[height=0.95em]{#1}}%
\parbox[c][10pt][t]{\wd0}{\box0}\endgroup
}

\DeclareRobustCommand{\inlinefigs}[1]{%
\begingroup
\setbox0=\hbox{\includegraphics[height=0.7em]{#1}}%
\parbox[c][8pt][t]{\wd0}{\box0}\endgroup
}

\newcommand{\Scenario}{\inlinefigs{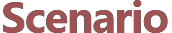}}
\newcommand{\Data}{\inlinefigs{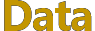}}
\newcommand{\Task}{\inlinefigs{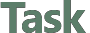}}
\newcommand{\EmbeddedVis}{\inlinefigs{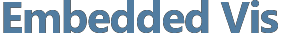}}
\newcommand{\Scenarios}{\inlinefigs{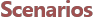} \hspace{-1.5mm}}
\newcommand{\Tasks}{\inlinefigs{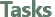} \hspace{-1.5mm}}
\newcommand{\Datap}{\inlinefigs{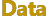} \hspace{-2mm}}

\newcommand{\ShotLabel}{\inlinefig{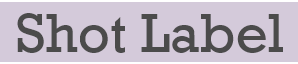}}

\newcommand{\Offense}{\inlinefig{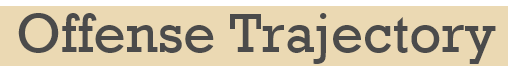}}

\newcommand{\Defense}{\inlinefig{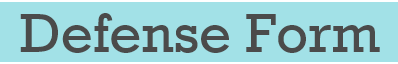}}

\newcommand{\ShotChart}{\inlinefig{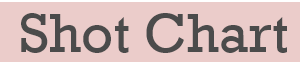}}

\newcommand{\TeamPanel}{\inlinefig{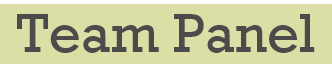}}

\newcommand{\Identify}{\inlinefig{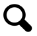}}
\newcommand{\Compare}{\inlinefig{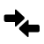}}
\newcommand{\Summarize}{\inlinefig{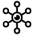}}

\newcommand{\Boxscore}{\inlinefig{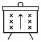}}
\newcommand{\Tracking}{\inlinefig{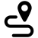}}
\newcommand{\Video}{\inlinefig{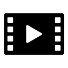}}

\newcommand{\Target}{\inlinefig{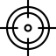}}
\newcommand{\Time}{\inlinefig{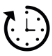}}
\newcommand{\Event}{\inlinefig{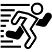}}

\newcommand{\Point}{\inlinefig{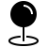}}
\newcommand{\Line}{\inlinefig{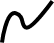}}
\newcommand{\Area}{\inlinefig{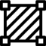}}
\newcommand{\Label}{\inlinefig{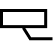}}
\newcommand{\Panel}{\inlinefig{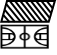}}

\definecolor{cb_orange}{rgb}{1.0,0.51,0.0}
\definecolor{cb_blue}{rgb}{0.22,0.49,0.72}
\definecolor{cb_green}{rgb}{0.3,0.67,0.29}
\definecolor{cb_red}{rgb}{0.89,0.1,0.11}
\definecolor{cb_purple}{rgb}{0.6, 0.31, 0.64}
\definecolor{cb_brown}{rgb}{0.6, 0.4, 0.2}
\definecolor{cb_crimson}{rgb}{0.86, 0.08, 0.24}

\newcommand{\re}[1]{{\textcolor{black}{#1}}}
\newcommand{\para}[1]{\vspace{1mm}\noindent\textbf{#1}}
\onlineid{1223}

\vgtccategory{Research}
\vgtcpapertype{Application/Design Study}

\title{The Quest for \OmniocularsTitle{}: Embedded Visualization for \\ Augmenting Basketball Game Viewing Experiences}

\author{Tica Lin, Zhutian Chen, Yalong Yang, Daniele Chiappalupi, Johanna Beyer, Hanspeter Pfister}

\authorfooter{
\item
 Tica Lin, Zhutian Chen, Johanna Beyer, and Hanspeter Pfister are with John A. Paulson
School of Engineering and Applied Sciences, Harvard University, Cambridge, MA. E-mail: \{mlin, ztchen, jbeyer, pfister\}@g.harvard.edu
 \item
Yalong Yang is with the Department of Computer Science,  Virginia Tech, Blacksburg, VA. E-mail: yalongyang@vt.edu
 \item
Daniele Chiappalupi is with the Department of Computer Science, ETH Zürich, Switzerland. E-mail: daniele.chiappalupi@inf.ethz.ch.
Work was done while Daniele Chiappalupi was an intern at Harvard University.
}

\abstract{Sports game data is becoming increasingly complex, often consisting of multivariate data such as player performance stats, historical team records, and athletes’ positional tracking information. While numerous visual analytics systems have been developed for sports analysts to derive insights, few tools target fans to improve their understanding and engagement of sports data during live games. By presenting extra data in the actual game views, embedded visualization has the potential to enhance fans’ game-viewing experience. However, little is known about how to design such kinds of visualizations embedded into live games.
In this work, we present a user-centered design study of developing interactive embedded visualizations for basketball fans to improve their live game-watching experiences. We first conducted a formative study to characterize basketball fans' in-game analysis behaviors and tasks. Based on our findings, we propose a design framework to inform the design of embedded visualizations based on specific data-seeking contexts. Following the design framework, we present five novel embedded visualization designs targeting five representative contexts identified by the fans, including shooting, offense, defense, player evaluation, and team comparison. We then developed Omnioculars, an interactive basketball game-viewing \re{prototype} that features the proposed embedded visualizations for fans' in-game data analysis. 
We evaluated Omnioculars in a simulated basketball game with basketball fans.
The study results suggest that our \re{design} supports personalized in-game data analysis and enhances game understanding and engagement.

} 


\CCScatlist{ 
 \CCScat{K.6.1}{Management of Computing and Information Systems}%
{Project and People Management}{Life Cycle};
 \CCScat{K.7.m}{The Computing Profession}{Miscellaneous}{Ethics}
}

\teaser{
 \hspace{-1cm}
 \includegraphics[width=1.1\linewidth]{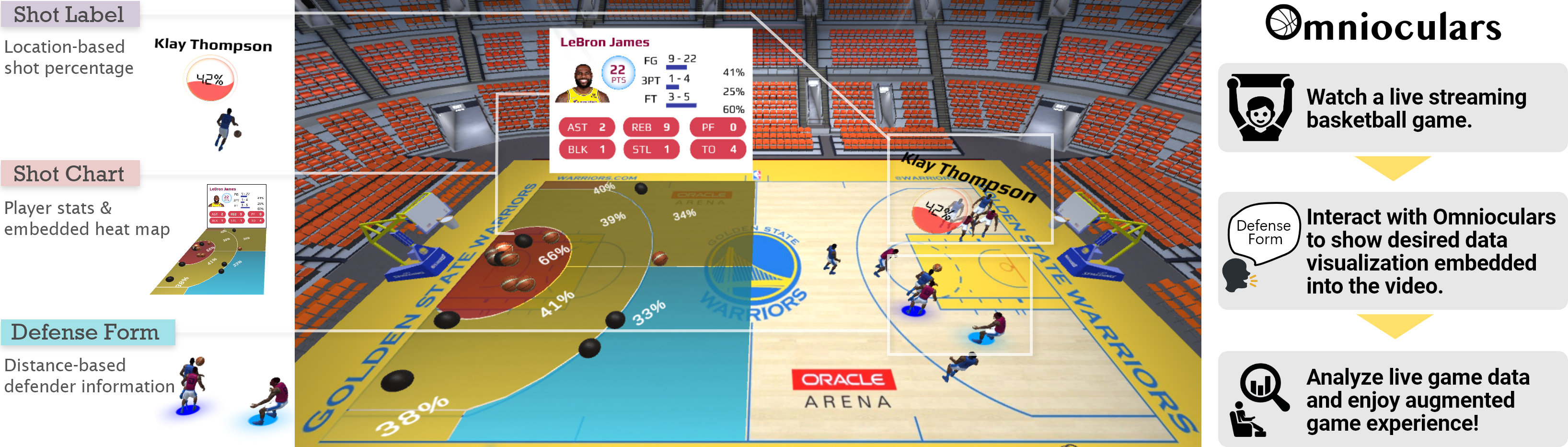}
 \caption{
 \re{Our Omnioculars prototype} supports a personalized and interactive game viewing experience with embedded visualizations for in-game analysis.  \re{We created a simulated basketball game environment to support our design probe of embedded visualizations.}
    }
    \label{fig:teaser}
}

\renewcommand{\manuscriptnotetxt}{}

\vgtcinsertpkg

\begin{document}



\firstsection{Introduction}
\maketitle

Sports broadcasting and streaming have seen exponential growth in recent years. 
Basketball, for example, is one of the most popular team sports worldwide. 
The NBA league alone averages about 1.4 million viewers per game on ESPN~\cite{espnnba}.
With the advent of novel sensing and image processing techniques,
an increasing amount of live data can now be collected in each NBA game, 
including scores, play-by-play data, and even player trajectories.
This data is released online in real-time and is widely used by sports analysts and coaches to measure the performance of teams, analyze the strengths and weaknesses of players, and make in-game decisions.
Overall, in-game sports data has gradually become indispensable
for professional basketball teams to develop winning strategies~\cite{tian2019use,mcintyre2016recognizing,76ers}. 
Several interactive sports analysis systems have been proposed to help experts consume and analyze in-game data~\cite{seidlBhostgustersRealtimeInteractive,losadaBKVizBasketballVisual2016,zhiGameViewsUnderstandingSupporting2019}. 
However, these systems are not yet readily available for sports fans. 
Subsequently, the information needs of a general audience during a game are largely unmet.

During a game, sports spectators and fans seek different data than expert analysts to facilitate their own game understanding and engagement.
After decades of development, scoreboards or data panels shown on TV are still the dominant method for spectators to obtain data during live games. 
However, presenting data in this way often does not fulfill the individual needs of the audience.
Consequently, perhaps the most common solution for spectators is to look up extra data on mobile devices or separate screens.
For example, basketball fans usually look up live box scores and play-by-play data during a game on the official NBA website~\cite{nba}, ESPN~\cite{espnnba}, or on the respective phone apps.
Such practices, however, distract from the game and limit the amount of in-game data analysis that would increase game understanding and engagement for fans and viewers.
Hence, spectators would greatly benefit from an effective way of analyzing in-game data and information that does not distract from the actual game.

Embedded visualization provides a promising opportunity to satisfy the in-game information needs of spectators, 
as it directly visualizes data within its physical context~\cite{willett2016embedded}. 
Thereby, embedded visualizations eliminate the need for context switching and reduce distractions when looking up data.
Recently, \re{some} commercial~\cite{secondspectrum} and research~\cite{chen2021} systems
have utilized embedded visualizations to augment sports videos (so-called \emph{augmented sports videos}),
providing extra information together with the video in a seamless and engaging manner.
However, all these systems are designed for experts to create augmented sports videos post-game 
and do not support custom visualizations for general audiences during a live game.
\re{In the live game, visualizations need to be simple and allow viewers to consume and analyze data on the fly,}
which can differ significantly from visualizations in post-game scenarios.
As such,
it is still an open question on how to best use embedded visualizations to satisfy the in-game \re{data} needs of a sports audience.

In this study, 
we aim to fill this gap by exploring the design space of embedded visualizations for live sports in-game data analysis.
We adopted a user-centered design approach throughout the study to answer three questions: 
First, we observe user needs and their workflow through a formative user study to answer \textit{``What data do fans desire in a live basketball game?''}.
Based on survey responses and in-depth user interviews with active basketball fans, 
we identified user needs in obtaining more advanced game data and control over their live game viewing experience. 
Next, we tackle the question of \textit{``What are the design considerations for embedded visualizations in live games, based on user needs?''}, 
by proposing a design framework 
for embedded sports data visualization based on 
\textit{scenario}, \textit{data}, \textit{task}, and \textit{embedded vis}.
We further identify five representative game contexts,
i.e., shooting, offense, defense, player performance, and team comparison,
and develop embedded visualizations for them.
To answer our final question of \textit{``How well do embedded visualizations help users enhance game understanding and engagement?''},
we developed \textbf{Omnioculars}, an embedded visualization \re{prototype} for live TV basketball game viewing \re{to support our design exploration}.
We took inspiration from Harry Potter's Omnioculars for Quidditch games~\cite{harrypotter}, which gives fans the ability to \emph{see everything} during a game.
We created
a simulated game environment in Unity3D~\cite{unity} based on NBA player tracking data to 
embed live visualizations on top of the simulated game.
We conducted a controlled user study with 16 active basketball fans using Omnioculars.
Users were able to generate distinct game insights with each embedded visualization 
and to utilize Omnioculars to personalize their in-game analysis in our simulated basketball game videos.

In summary, our contribution consists of the first design study on embedded visualization for in-game data analysis of basketball fans.
Second, we propose a design framework to support embedded visualization designs based on game contexts.
Third, we built a simulated game-watching environment and implemented Omnioculars, 
a novel design prototype for showing five embedded visualizations in live basketball game videos based on our design framework.
Finally, we evaluate the merits and limitations of embedded visualization for in-game analysis with basketball fans in a user study.
\section{Related Work}

\para{Visual Analytics for Sports.}
Advances in sensors and computer vision techniques have led to the collection of more fine-grained, high-quality sports data.
Perin et al.~\cite{Perin2018-jh} categorize sports data in visualizations into box score data, tracking data, and metadata. 
Given the highly competitive nature of sports, 
numerous visual analytics systems have been developed to allow domain experts to analyze complex sports data for discovering winning strategies.
For example, SoccerStories~\cite{perinSoccerStoriesKickoffVisual2013} employs a series of soccer-related visualizations, such as in-court trajectories and heatmaps, 
to analyze spatio-temporal data collected in soccer games.
BKViz~\cite{losadaBKVizBasketballVisual2016}, an interactive data exploration tool for basketball games, focuses on the analysis of play-by-play data.
Fu and Stasko~\cite{fu2022supporting} developed interactive visualization systems specifically targeting NBA journalists.
Some sports analytics approaches also leverage machine learning models to predict and extract patterns from the data~\cite{tian2019use,mcintyre2016recognizing,nistala2018using}.
Recently,  with the proliferation of low-cost immersive devices such as Virtual and Augmented Reality (VR/AR),
immersive sports analytics systems have been gaining traction.
Compared to traditional desktop systems, immersive sports analytics systems provide unique benefits, including increased spatial understanding, rich embodied interaction, peripheral awareness, and a large information space~\cite{marriott2018immersive,lin2020sportsxr}.
ShuttleSpace~\cite{ye2020shuttlespace}, a VR-based visual analytics system for badminton data,
enables users to analyze shuttle trajectories from a player's first-person view.
Following their work, 
TIVEE~\cite{chuTIVEEVisualExploration2022} allows experts to explore a large scale of badminton trajectories from both first and third-person views in VR.
While these systems have demonstrated their effectiveness in analyzing sports data, 
most of them are designed for post-game analysis by domain experts, 
leading to a high entry barrier and steep learning curve.
On the other hand, 
compared to post-game analysis systems, 
in-game systems usually can be used by both domain experts and fans to complete lightweight analysis.
However, only a few approaches focus on in-game analysis.
GameViews~\cite{zhiGameViewsUnderstandingSupporting2019}, for example, supports data-driven sports storytelling using in-game data for sportswriters and fans.
Yet, GameViews mainly uses simple visualizations separated from the videos, 
forcing users to switch contexts when watching a game.
We draw on this line of research to develop an in-game analysis tool for \emph{basketball fans},
which features embedded visualizations
to enable lightweight analytics while retaining an engaging game experience.

\para{Embedded Visualization in Sports Videos.}
Sports data, generally speaking, is inherently associated with its physical environment (e.g., a basketball court).
Visualizing sports data within this physical environment can significantly facilitate the understanding and analysis of the data.
Thus, numerous works have explored visualizing sports data in static court diagrams.
For example, CourtVision~\cite{goldsberry2012courtvision} uses point-based visualizations to present the shot distribution and density for NBA data.
Sacha et al.~\cite{sacha2017dynamic} introduced a trajectory-based visualization system that allows users 
to interactively explore the moving trajectories of soccer players.
To improve the analysis of soccer data, 
Stein et al.~\cite{DBLP:journals/cga/SteinJBSSGCK16} developed a region-based visualization that shows the interaction space of each soccer player.
In recent years, with the rapid development of computer vision and image processing techniques,
visualization researchers have started to directly embed their visualizations into sports videos.
For instance, 
Stein et al.~\cite{steinBringItPitch2018} developed a system to assist soccer game analysis, which automatically extracts and visualizes data from and in soccer videos.
Stein et al.~\cite{stein2018} then extended their work by semi-automatically selecting the presented data based on the game's semantic context.
However, these works mainly target experts for analytic purposes and do not fulfill general spectators' needs. 
More recently, Chen et al.~\cite{chen2021} explored the design space of augmented sports videos and presented VisCommentator, a rapid prototyping tool for authoring augmented racket-based sports videos.
In summary, current work in augmented sports videos focuses on domain experts' analytic and authoring needs. 
Embedded visualizations designed for sports fans are still under-explored. 
In this work, we aim to understand fans' in-game analysis needs and propose a structured design framework for embedded visualization in basketball videos.

\para{Personalized Game Viewing Systems.}
When watching a sports game, 
spectators often look up extra information to satisfy their curiosity, 
deepen their understanding of the game, or generate game insights \re{(as found in our formative study in Sec~\ref{sec:design-requirement-analysis})}.
These information needs vary from person to person, leading to a great demand for personalized game viewing systems.
The most common way fans access additional data is using mobile devices to search the internet when watching the game.
However, this approach inevitably incurs context-switching costs
and significantly impedes an engaging watching experience. 
To overcome this issue, HCI and Computer Graphics researchers have studied interactive game viewing systems 
that enable spectators to look up data while still focusing on the game. 
For instance, Gamebot~\cite{zhiGameBotVisualizationaugmentedChatbot2020} uses a chatbot interface on a mobile app for fans to request and consume game data visualization. 
ARSpectator~\cite{zollmannARSpectatorExploringAugmented2019} uses mobile AR devices to overlay basic game data onto the scene for on-site spectators.
Despite the focus on interaction techniques to access game data, 
the existing work did not evaluate the presentation of game data to facilitate game understanding and engagement.
On the other hand,
E-Sports, such as Defense of the Ancients2~\cite{dota} (Dota2) and League of Legends~\cite{lol} (LoL), are ahead of most traditional sports in providing personalized game viewing experiences due to their interactive, online nature.
In Dota2, for example, the audience can interact with the viewing system to inspect the gaming data (\emph{e.g.}, gold over time) of individual players or the whole team.
Nevertheless, most of these systems only present the data in side-by-side settings, in which the visualizations are displayed in panels separated from the game view.
We envision that embedded visualizations have the potential to seamlessly blend the extra information with the game view, providing a natural, intuitive, and engaging watching experience.
Our work takes a first step towards designing and evaluating personalized game viewing systems with embedded visualizations for traditional sports.

\begin{figure}[t!]
    \centering
    \includegraphics[width=\linewidth]{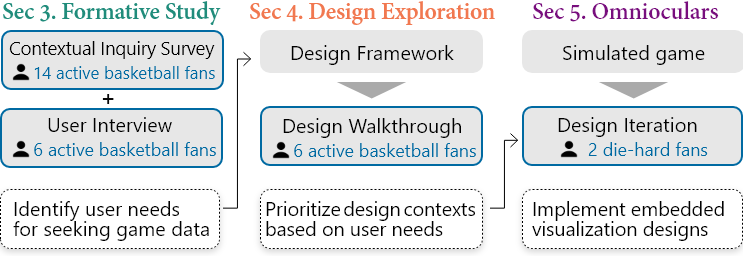}
    \vspace{-6mm}
    \caption{We adopted a user-centered design process throughout the study. We elicited user needs through a survey and interviews, and iterated our designs based on user feedback. } 
    \label{fig:process}
    \vspace{-4mm}
\end{figure}

\section{Formative Study with Basketball Fans}
\label{sec:design-requirement-analysis}

Section~\ref{sec:design-requirement-analysis}, \ref{sec:design-space-exploration}, and \ref{sec:omnioculars} present our user-centered design process towards designing an interactive basketball game-viewing system with embedded visualizations (\autoref{fig:process}).
In this section, we detail a mixed-method formative study using an online survey and
interviews with active basketball fans to better understand the general workflow, pain points, and best practices of fans to acquire desired data during a live game.



\subsection{Procedures}
\label{sec:contextual-inquiry}


We first used an online survey to collect initial feedback on basketball fans' data needs while watching live games.
We analyzed survey responses to understand fans' in-game data analysis behaviors
and to inform the questions of the follow-up interviews (\autoref{fig:process}).

\para{Online Survey.}
The game-watching experiences survey \re{(Appendix A)}
asked about fandom levels (\emph{novice}, \emph{casual}, \emph{engaged}, and \emph{die-hard} fans), game watching frequency, \re{and in-game data analysis records}.
We asked respondents to watch a live basketball game 
and record three moments when they wanted to look up additional data. 
We advertised the survey using university mailing lists
and only collected responses from casual fans and above.
In total, we gathered 14 responses (R1-14; M = 10, F = 4; Age: 18 - 45) \re{and 40 records of their data-seeking moments when watching a live game}.
All respondents are NBA fans (4 casual, 6 engaged, and 4 die-hard fans).

\para{In-depth Interview.}
We then interviewed six fans from the survey respondents (I1-6; 4 die-hard and 2 engaged fans) who expressed interest in the survey. 
Among them, five have been NBA basketball fans for over ten years.
We conducted the interviews remotely using Zoom in two parts (Appendix B).
We first asked respondents about their overall game viewing experiences 
and data analysis behaviors in recent games they watched, 
including occasions they looked at data and whether the data was helpful and engaging.
We then asked them to watch three video clips selected from a 9-minute post-game highlight video of an NBA regular-season game\footnote{Golden State Warriors vs. Cleveland Cavaliers, 2015} and think aloud about what game insights and data they observed and wanted to see.
Each clip consisted of two to four plays for about 30 seconds and collectively covered a diverse range of gameplays, 
including offense strategy, a player's highlight (a layup with a foul), defense performance (a block), or clutch time.
The clips also included video effects to augment certain gameplay, such as replays from different angles, showing a player's box score on the screen, etc.
The interviewees contributed 63 in-game data analysis records on the three clips. \re{No further interviews were conducted as saturation was met in the analysis.}


\para{Analysis.}
We performed a reflexive thematic analysis~\cite{braun2019reflecting} on the survey and interview data. 
Three authors began coding separately on the 40 survey records to create sets of plausible codes, 
and through later discussion and affinity diagramming with another author,
agreed on a single coding scheme.
Then the lead author finished the coding on the remaining records collected from the interviews.


\subsection{Findings and Discussions}
\label{ssec:survey-interview-findings}

We present findings on fans' motivation, workflow, and pain points of performing in-game data analysis.

\subsubsection{Practices of Watching Live Basketball Games}

Overall, most of the respondents watched NBA games at least once per week (4 watched 3 games per week, 9 watched 1-3 games per week, 1 watched less).
12 of them followed a team or a player in the NBA league.
All six interviewees watched 1-3 NBA games per week, 
mainly to follow their favorite players. 
As for the game watching media,
the interviewees explained that while watching games on-site is engaging and exciting for all fans, 
they mostly watch live streams for convenience and economic reasons. 
When asked to compare their game-viewing experience between live streams and highlight videos, 
interviewees expressed that 
they preferred to watch live streams over highlights. 
One key reason is that highlight videos do not present the flow of the game. 
Although highlight videos usually provide additional visual effects, 
such as play breakdown or embedded data overlays, 
fans cannot observe how the game developed to the current moment and feel \textit{``it shows no rhythm of the game''} (I4). 
In summary, compared to on-site watching and highlight videos,
live streaming is the favorite game watching medium among fans \re{we surveyed}.


\subsubsection{Seeking Data in Various Contexts During Live Games}
\label{ssec:data_context}

Seeking data while watching a live game \re{was a common behavior among fans we surveyed}.
While watching a game,
all \re{14} respondents in the survey expressed interest in looking up game data; 
all six interviewed fans confirmed that they 
always look up game stats to complement the viewing experience.
Based on our analysis of the records, 
a context in which a fan seeks data while watching a live game
can be characterized by a \Scenario{}, \Data{}, and \Task{}.
Specifically, 
in different scenarios (When), 
fans seek different data (What) to accomplish different tasks (Why):

\para{Fans seek data in three typical \Scenarios{} (When)}.
First, fans often look for data at \emph{\textbf{specific game times}}, such as before the game, during halftime, or in the last quarter.
These specific game times are defined based on the rules of the sport, like basketball, and thus usually have special meanings.
Second, fans seek data when observing \emph{\textbf{a game event}}.
These events include seeing a game stat, a game action, or the appearance of a new player, \textit{``D. Rose made an and-1 play in the 2nd quarter''} (R8).
Third, when \emph{\textbf{multiple repetitive game events}} happen,
fans sometimes want to seek data to evaluate a player's or team's performance related to this specific type of game event, \textit{``Thunder is missing a lot of 3 pointers''} (R5).

\para{Fans seek two types of \Datap{}  (What).}
Based on the survey and interviews, the data sought by the fans can be divided into two types, inspired by Perin et al.~\cite{Perin2018-jh}.
First, \emph{\textbf{game data}} are directly collected from games, 
including box score, \textit{``foul number of Embiid''} (R4),  
tracking data, \textit{``the usage of this particular play''} (R9), 
and video data. 
The game data sought by the fans can range from a single game, to multiple games, or an entire career.
Second, \emph{\textbf{metadata}} includes data beyond games, such as ranking, player background, and news. 

\para{Fans seek data to finish three \Tasks{}  (Why)},
ultimately generating insights to deepen their understanding of the game.
Inspired by  Brehmer and Munzner~\cite{brehmer2013multi,munzner2014visualization},
we summarized three typical tasks that the fans perform with data during a live game.
First, fans seek data to \emph{\textbf{identify}} the performance of players or teams.
For example, a respondent replied, \textit{``I was wondering whether he was getting a triple-double''} (R4). 
In addition to looking at the data of a specific player, fans are also interested in
the data of the whole team, such as the moving trajectories of a team to 
``\emph{understand the offensive and defensive strategies}'' (I3).
Second, fans often need data to \emph{\textbf{compare}} the in-game performance of a player or team 
to others' in-game performance or to their historical performance.
For example, a respondent replied, 
\textit{``I want to know how many assists our team makes compared to other teams''} (R5).
Five interviewees also provided that they wanted to compare a player's in-game performance
to their average performance to evaluate the consistency.
Third, fans usually seek data to \emph{\textbf{summarize}} the performance of players or teams.
Compared to the other two tasks, as pointed out by Brehmer and Munzner~\cite{brehmer2013multi,munzner2014visualization},
the summarize task often involves a deeper analysis of the data, such as overviewing the game and observing patterns, 
deriving the cause of the performance, 
and speculating on possible outcomes. 
Representative responses include \textit{``to have a quick understanding of both teams so I know what to expect in the match''} (R10) (overview \& possible outcome) and 
\textit{``to understand what causes the score difference between two teams''} (R3) (cause of the performance). 

\subsubsection{Pain Points of Seeking Data in Watching Live Streams}
While seeking data \re{was} common among basketball fans,
the data acquisition process in practice remains hindered by current technologies.
According to the survey, the main method of acquiring data is through websites or mobile apps (30 out of 40).
In the interviews,
the fans further detailed that they access websites (ESPN~\cite{espnnba}, NBA~\cite{nba}, or Reddit~\cite{reddit}) 
on the phone or on a separate screen to regularly track box score data or team standing. 
Such practices reflect several issues in the live game-watching experience:

\para{Data provided by live streams lack diversity.}
The interviewees agreed that 
most information available on live streams, such as game stats and timers, 
was useful.
However, they also confirmed that the information provided by the live streams is not diverse and sufficient enough, e.g., \textit{``usually they are not sufficient, only showing a matrix of all player names and scores (I2)''}.
Some information that the interviewees would love to have is usually absent in live streams, 
such as seeing all box scores and lineups information, or comparing shooting performance by zones.
This is the main reason that fans search websites to obtain extra data while watching live games.

\para{Data provided by live streams lack controllability.}
Interestingly, although live streams can provide some basic box score data (like game stats and timers),
the interviewees still will check this data on the website regularly, 
\textit{``Right now, on a separate screen, I can look at the full box scores to get an updated impression of the players''} (I4).
This practice reflects that when watching live games,
fans need not only sufficient data but also the ability to control when and what data to show.
This finding also links to our characterization that users seek different data in various contexts (Sec.~\ref{ssec:data_context}).

\para{Data provided by websites are separated from the game videos.}
Unsurprisingly, the interviewees complained that looking up data on the phone or a separate screen reduces the engagement of the game watching experience as it introduces numerous context switches between the game videos and websites.
For example, \textit{``in a live game there’s a lot going on and likely I missed some events [when looking up data]''} (I1).
Thus, interviewees would like a more intuitive method to present the data directly within the game videos.

\subsection{Summary}

Our \re{formative study} found that live streams \re{were} the favorite watching medium for basketball fans \re{we interviewed}.
While watching live games,
fans seek different types of data in different scenarios for different tasks. 
However, current live streams \re{fall short in fulfilling fans' diverse data needs,} 
forcing fans to look up data on separate screens (phones or monitors),
which reduces the engagement in watching games.
To tackle these issues,
we propose to \re{embed visualizations into live game videos to present data in-situ and support in-game data analysis.}
\section{Design Space Exploration}
\label{sec:design-space-exploration}

\begin{figure}[t!]
    \centering
    \includegraphics[width=\linewidth]{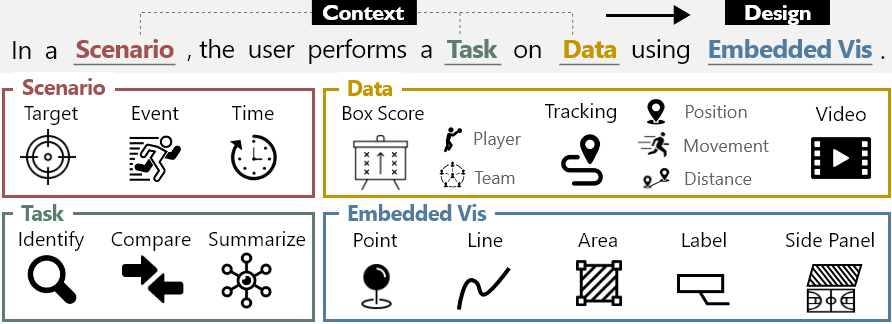}
    \caption{Design framework for context-driven embedded visualizations.} 
    \label{fig:design_framework}
    \vspace{-3mm}
\end{figure}

To answer our second research question of how to design embedded visualizations for fans' in-game data analysis, 
we first developed a design framework based on the formative study findings (\autoref{fig:design_framework}).
With this framework, we then identified 20 contexts from the \re{records of fans' in-game data seeking moments}. By walking through these design contexts using design mockups with the six interviewees,
we prioritized five contexts to be implemented in our \re{prototype}. 


\subsection{Design Framework}

Based on the data-seeking contexts we summarized in Sec.~\ref{ssec:data_context},
we propose a context-driven design framework for embedded visualizations for in-game data analysis (\autoref{fig:design_framework}). 
The framework scaffolds the relationship between four design aspects -- 
``\emph{In a certain \Scenario{}, the user performs a \Task{} on \Data{} using \EmbeddedVis{}}'':

\begin{itemize}
    \item \Scenario{} describes the background of fans' data analysis behaviors.
    Three design elements can be used to describe a scenario that drives a fan's data need:
    \Target{} \emph{target} (team or player),
    \Event{} \emph{event} (any game event that is happening), 
    and \Time{} \emph{time} (a specific moment or a duration). 
    For example, \textit{``When Joel Embiid fouled''} describes a player (target) with a foul (event) at the current time (time);
    \textit{``When we have been making strides with offense for a streak of time in the game''} describes a team (target) make shots (event) over a certain period (time). 

    \item \Data{} that is of interest to fans can be collected in-game or beyond games.
    For embedded visualizations, we focus on in-game data and exclude metadata.
    Thus, the data types to be considered include 
    \Boxscore{} \emph{box scores} (game summary scores), 
    \Tracking{} \emph{tracking data} (like player positions), 
    and \Video{} \emph{video}.
    Different data can be further combined to specify advanced stats, 
    such as \textit{``the lineup's plus-minus''} or \textit{``the shot percentage of \#30 when there is no defender in the front.''}

    \item A \Task{} is performed on game data by fans to generate insights about players or teams,
    including \Identify{} \emph{identify} (on a player or team), 
    \Compare{} \emph{compare} (between players or teams), 
    and \Summarize  \emph{summarize} (among players or teams).
    
    \item \EmbeddedVis{} can present in-game data directly in videos to support fans performing in-game data analysis naturally.
    Prior research~\cite{chen2021} systematically studied the design space of augmented sports videos
    and summarized the embedded visualizations used in augmented videos to present sports data.
    We follow their design space with a focus on embedded visualizations for basketball,
    and identified five relevant graphical marks and basic visual elements:
    \Point{} \emph{point}, 
    \Line{} \emph{line}, 
    \Area{} \emph{area}, 
    \Label{} \emph{label}, 
    and \Panel{} \emph{side panel}. 
    These basic visuals can be composited to construct embedded visualizations in basketball game viewing.

\end{itemize}

This design framework helps designers 
identify fans' in-game data analysis contexts 
by specifying the \Scenario{}, \Data{}, and \Task{}.
Such a context captures the fan's specific requirements for in-game data analysis at a given moment.
Once a context is identified, the designer can then focus on the design of embedded visualizations 
by composing the five basic visuals,
thereby ensuring the embedded visualizations are developed based on and can satisfy the fan's analysis needs.

\begin{figure}[t!]
    \centering
    \includegraphics[width=\linewidth]{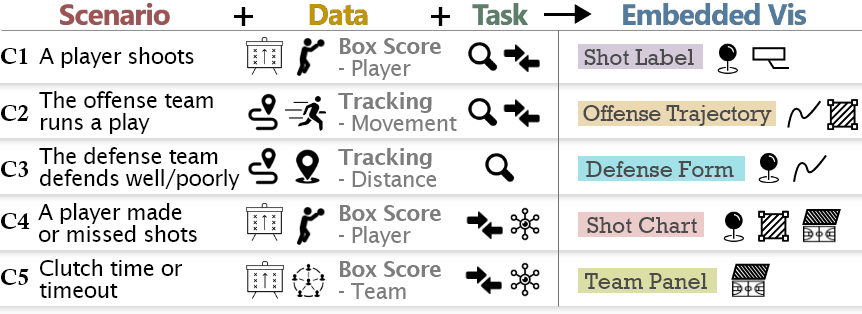}
    \caption{We developed five embedded visualization designs based on our design framework and the five prioritized contexts in Sec.~\ref{sec:scenario-walkthrough}.} 
    \label{fig:design_context}
    \vspace{-3mm}
\end{figure}

\subsection{Design Context Walkthrough}
\label{sec:scenario-walkthrough}

Following our design framework,
we identified 20 design contexts from 103 
\re{records of in-game data seeking moments}
in the formative study 
\re{by grouping similar records and summarizing representative contexts}.
We created 20 design mockups to demonstrate the contexts (Appendix C) and 
walked through them with 
the six fans in follow-up interviews.
Based on their ratings of importance, we prioritized five contexts that drove the design of five embedded visualizations in our \re{prototype} (Fig.~\ref{fig:design_context}).

\begin{itemize}
    \item[\textbf{\texttt{C1}}] \texttt{\textbf{Shooting}} -- A player shoots (\Scenario{}), 
    and the fan wants to identify and compare (\Task{} \Identify{}\Compare{}) the player's box score (\Data{}). 
    Fans found this context useful and suggested avoiding information overload by only showing the most important stat based on player and location. 

    \item[\texttt{\textbf{C2}}] \texttt{\textbf{Offense}} -- The offense team runs a play (\Scenario{}), 
    and the fan wants to identify and compare (\Task{} \Identify{}\Compare{}) the players' movement (\Data{}). 
    Fans commented that this context is very useful to follow the ball and understand the play better, especially for tracking off-ball players and identifying open shot opportunities. 

    \item[\texttt{\textbf{C3}}] \texttt{\textbf{Defense}} -- The defense team defends well/poorly (\Scenario{}), 
    and the fan wants to identify (\Task{} \Identify{}) the defense scheme (\Data{}). 
    Fans commented that 
    this context helps identify the effective defenders and track how defensive focus is changing. 

    \item[\texttt{\textbf{C4}}] \texttt{\textbf{Player Performance}} -- A player has made/missed shots (\Scenario{}), 
    and the fan wants to compare and summarize (\Task{} \Compare{}\Summarize{}) the player's shot performance by zones (\Data{}).
    Fans found this context helpful for evaluating a player's game performance in more detail with direct reference to their average performance.

    \item[\texttt{\textbf{C5}}] \texttt{\textbf{Team Comparison}} -- At the clutch time (\Scenario{}), 
    the fan wants to compare and summarize (\Task{} \Compare{}\Summarize{}) team stats (\Data{}). 
    Fans commented instant access to this data is very convenient during the game break. 
\end{itemize}




\section{\Omnioculars{}}
\label{sec:omnioculars}

Building upon our design space exploration in Sec.~\ref{sec:design-space-exploration},
we present Omnioculars, 
an interactive embedded visualization \re{prototype} for live game data analysis for fans.
Omnioculars consists of three main components:
a game simulator, embedded visualizations, and interactions.


\subsection{Game Simulator}
\label{sec:data-source}

Ideally, Omnioculars should be built on top of live stream TVs 
as a third-party game viewing system
to enhance the real-world watching experience.
However, we decided to develop 
Omnioculars on top of a game simulator instead of real-world broadcasting videos, considering two aspects:
On the one hand,
implementing Omnioculars based on publicly available broadcasting videos is extremely difficult.
First, the data that can be extracted from broadcasting videos is limited
due to the reduced frame rate and low resolution.
Second, rendering embedded visualizations into broadcasting videos is challenging 
as the camera parameters of these videos are usually not made public.
Third, broadcasting videos often switch between cameras and insert replays,
reducing the controllability of Omnioculars.
On the other hand,
our goal is not to develop Omnioculars as a fully functional system used in real-world scenarios
but as a design probe~\cite{DBLP:conf/chi/HutchinsonMWBDPBCEHRE03} to explore the notion of using embedded visualizations to improve game viewing experiences.
To gain feedback on our design without being bogged down by the aforementioned engineering challenges,
we use a game simulator as the system's basis, 
thereby obtaining richer data, better rendering results, and controllability.
Please note that the aforementioned challenges can be tackled 
once the source video streams 
and sensing data from the live game collected by the TV companies and NBA league
are available.

{\renewcommand{\arraystretch}{1.2}
\begin{table}
    \centering
    \caption{The required data for the game simulator.}
    \label{tab:data}
    \vspace{-0.5em}
    \begin{tabularx}{\columnwidth}{p{2cm} p{5.7cm}}
    \toprule
    
    \textbf{Type} & \textbf{Description} \\ \midrule
    Tracking Data & The players and the ball's position over time. \\ \hline
    Event Data & The game events, each event is described by its time, player, action, and outcome. \\ \hline
    Shot Locations & The position of each shot event. \\ \hline
    Box Score & Team and player statistics of the game. \\ \hline
    Player Zoned Shot  \%  & Player's detailed shot percentage by shot locations grouped into zones.   \\ \hline
    League Average Zoned Shot \% & The NBA league's average shot percentage by shot locations grouped into zones.\\
	\bottomrule
\end{tabularx}
\vspace{-4mm}
\end{table}
}

The game simulator is implemented using Unity3D~\cite{unity},
with a 3D basketball arena and 10 player models (\autoref{fig:teaser}).
To simulate a basketball game, we adopt a data-driven strategy.
Specifically, the simulator takes a set of spatio-temporal data as the input
and outputs a virtual basketball game 
by moving the players and triggering events based on the data.
Table~\ref{tab:data} shows the required data in our current implementation.
Overall, 
as suggested in the user study (Sec.~\ref{sec:game-simulation}),
the simulated game was perceived as comparable to an actual game in terms of acquiring game insights by the users.

\subsection{Embedded Visualization Designs}
We iterated over each of our visualization designs with two fans we interviewed.
Both were die-hard fans with substantial basketball knowledge and game viewing experiences. Therefore, we considered them to be domain experts and asked for feedback on our visualization designs.

\subsubsection{Shot Label} 
\begin{figure}[h]
  \centering
  \includegraphics[width=\linewidth]{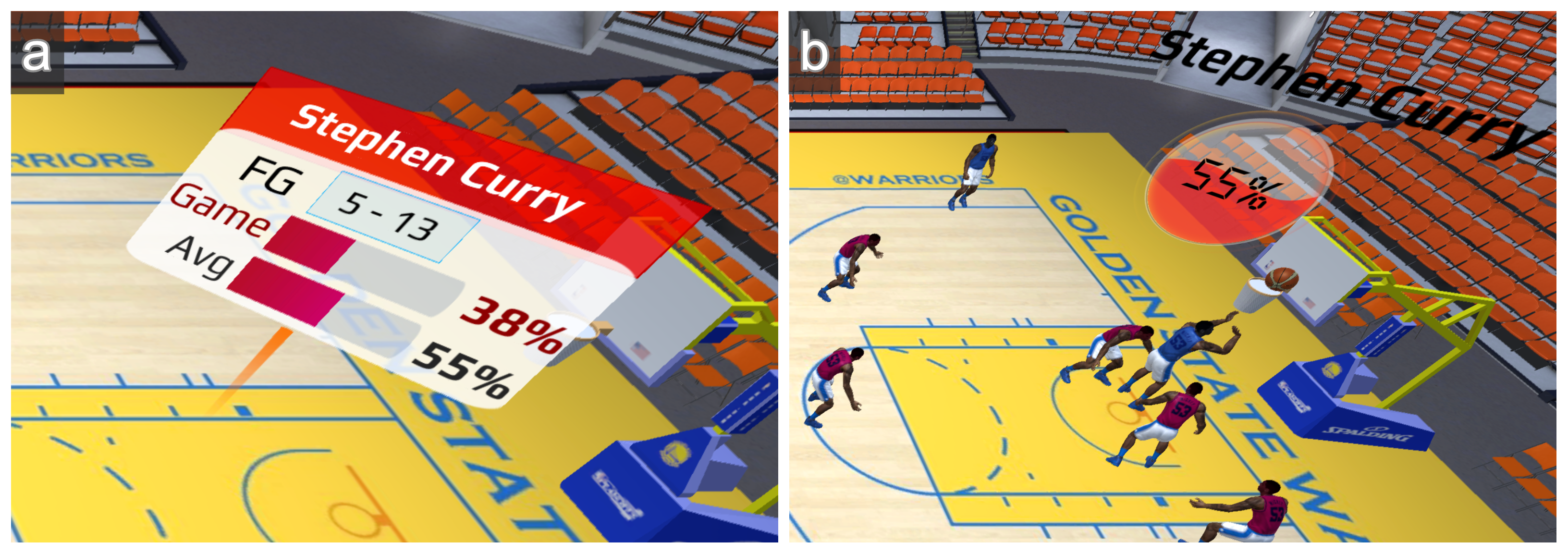}
  \vspace{-5mm}
  \caption{Shot Label shows (a) static outcome and (b) dynamic percentage.}
  \label{fig:ev1}
\end{figure}

\noindent
In \texttt{C1 Shooting}, 
we identified that fans want to see the shot performance of the shooter before and after shooting to predict and evaluate the shot outcome,
and compare the shooter's shot percentages to their average or others' performance. 
We propose a \ShotLabel{} that consists of two components:
1) a static label with the shooter's shot outcome at the shot location (\autoref{fig:ev1}a)
and 2) a dynamic label showing the player's shot percentage based at their current location (\autoref{fig:ev1}b).

The static 
label is shown at the shot location after a player shoots and is visible for 5 seconds and
presents the shot outcome, the shooter's game, and the average shot percentage with bar charts. 
This allows fans to get a quick comparison of the shooter's own in-game performance without interfering with the continuing game. 

The dynamic label shows the shooter's average shot percentage based on the location above their head and moves with the player. 
This supports comparing the player's shot performance among their team and instantly evaluating the player's shot opportunity based on location. 
The dynamic shot percentage is obtained from one of the seven shot chart zones (see Fig.~\ref{fig:ev4}). 
To support direct visual comparison, 
the label is color-coded to compare the player's shot percentage to the league's average, 
colored in red (hot) if above and in blue (cold) if below. 
The dynamic label supports fans to compare the shooter's shot percentage to other players, 
such as evaluating if the shooter is shooting from their hot spot, and the quality of shot selection.

Alternatively, our original design had a label pop up right after a shot
and move with the shooter for a few seconds, showing the score and shot percentage of the shooter.
However, fans' feedback suggested that dynamic labels with dense information are difficult to read and showing a player's shot outcome as the player moves into the next play does not help with game analysis. 
Thus, we abandoned this design.


\subsubsection{Offense Trajectory}

\setlength{\intextsep}{0pt}%
\begin{wrapfigure}{R}{0.22\textwidth}
	\centering
	\includegraphics[width=0.22\textwidth]{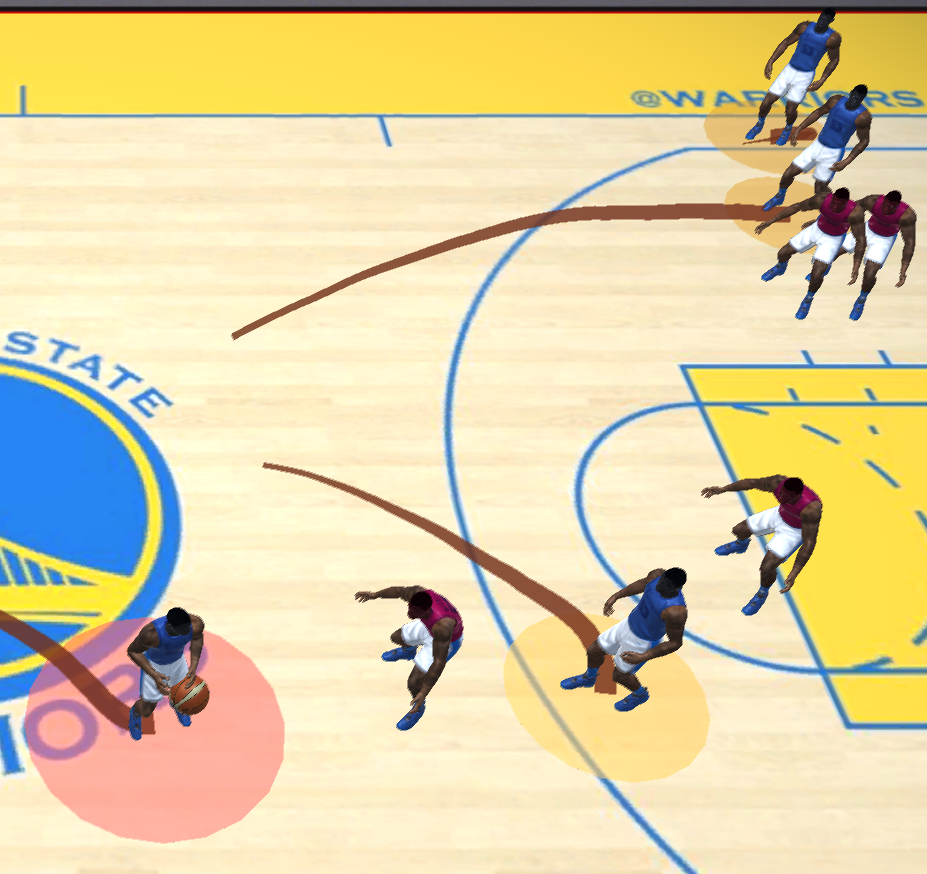}
	 \vspace{-4mm}
	\caption{Offense Trajectory}
	 \vspace{-2mm}
	\label{fig:ev2}
\end{wrapfigure}

In \texttt{C2 Offense},
fans were eager to use visualization to track the playset and spot open shot opportunities from the off-ball players.
Our \Offense{} design (\autoref{fig:ev2}) shows the offense players' trails and highlights the open space around an offense player with a circle underneath the player (area). 
The radius of the open space is drawn based on the player's distance to the closest defender and changes dynamically to reflect the interaction between the offense and defense team. 
We also color the ball handler's open space in red to guide the viewer's attention. 
Fans can identify the play run by the offense team through the trails, 
and easily tell the speed and rotation of the players. 
The open space circles reveal how the defense team reacts to the play and help fans immediately spot open shot opportunities (larger circles). 

Before finalizing the design, we iterated on the design 
by encoding other useful tracking data, 
such as speed measurement, player role, and distance to other players. 
User feedback suggested that encoding speed and player roles does not provide much value as a player's speed changes constantly 
and does not represent the actual moves of the player. 
The differentiation of player roles also became less prominent in modern basketball games. 
However, encoding the distance to the defense player was deemed very useful for fans to track the execution of the play.
We thus kept this encoding.


\subsubsection{Defense Form}
\vspace{1mm}

\begin{figure}[h]
  \centering
  \includegraphics[width=\linewidth]{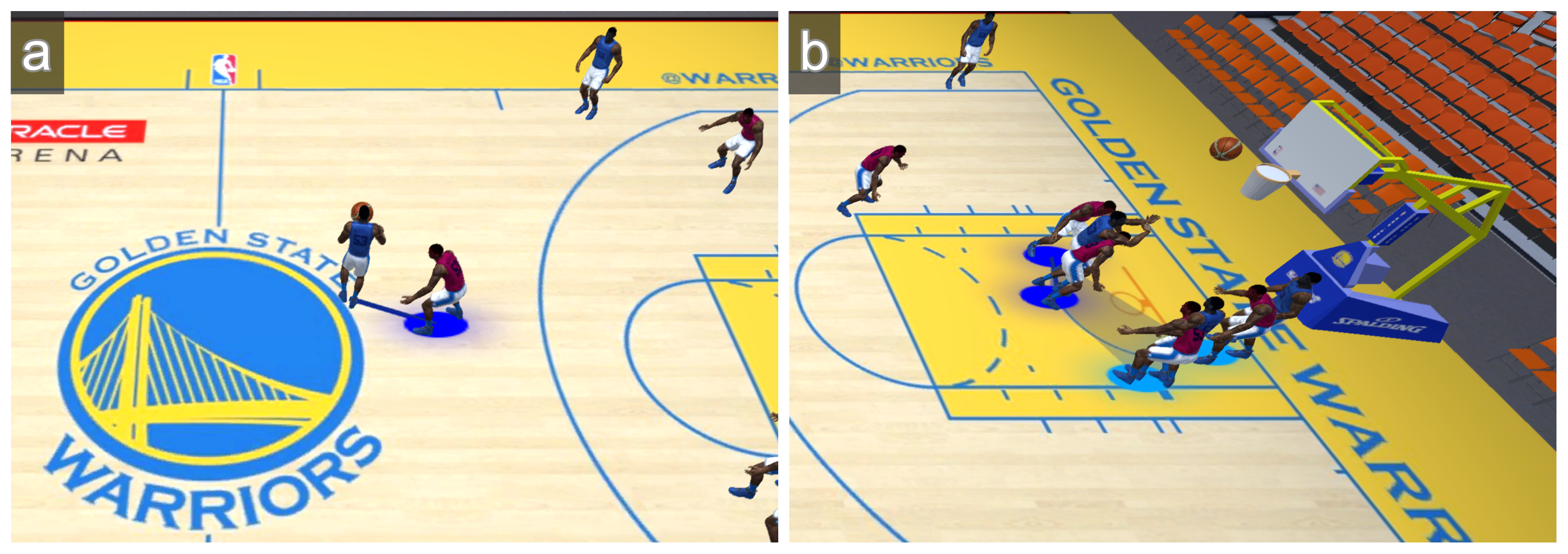}
  \vspace{-5mm}
  \caption{Defense Form shows (a) ball defender and (b) defensive focus.}
  \label{fig:ev3}
\end{figure}

\noindent
In \texttt{C3 Defense}, 
fans wanted to identify the defense scheme and key defenders, 
and track the change of defenders when a switch happens. 
The \Defense{} 
highlights the key defenders by tracking the distance between the defenders and the ball handler 
(\autoref{fig:ev3}a). 
When a defender is within $6 \ft$ of the ball, that player is marked as the ball defender (dark blue) and connected to the ball handler with a line. 
The defender is marked as a helper (blue) if within $12 \ft$ of the ball. 
When multiple defenders are identified as key defenders on the strong side (the ball handler's side of the court), the enclosed area is colored to highlight the region of defensive focus (\autoref{fig:ev3}b). 
Fans can evaluate the defense scheme 
from these visual marks without being overwhelmed.

An alternative design was to highlight the position of all defensive players (point) 
and color the enclosed region (area) to achieve the goal of tracking the defense scheme and transformation. 
However, this design did not specify the key defenders, 
including the ball defender and helpers on the strong side. 
Furthermore, the large enclosed area can cause visual clutter,
and does not represent the actual defense focus, 
as the further away the defense players are from each other, the less of a threat they are. 
Instead of highlighting all defensive players, 
our final design focuses on
drawing user attention to the effective defenders and showing the defensive focus of the enclosed area between key defenders.     


\subsubsection{Shot Chart}

In \texttt{C4 Player Performance}, 
fans valued the benefit of using embedded visualization to compare the shooter's shot performance in game to their own seasonal average in more detail. 
\begin{wrapfigure}{r}{0.21\textwidth}
	\centering
	\includegraphics[width=0.21\textwidth]{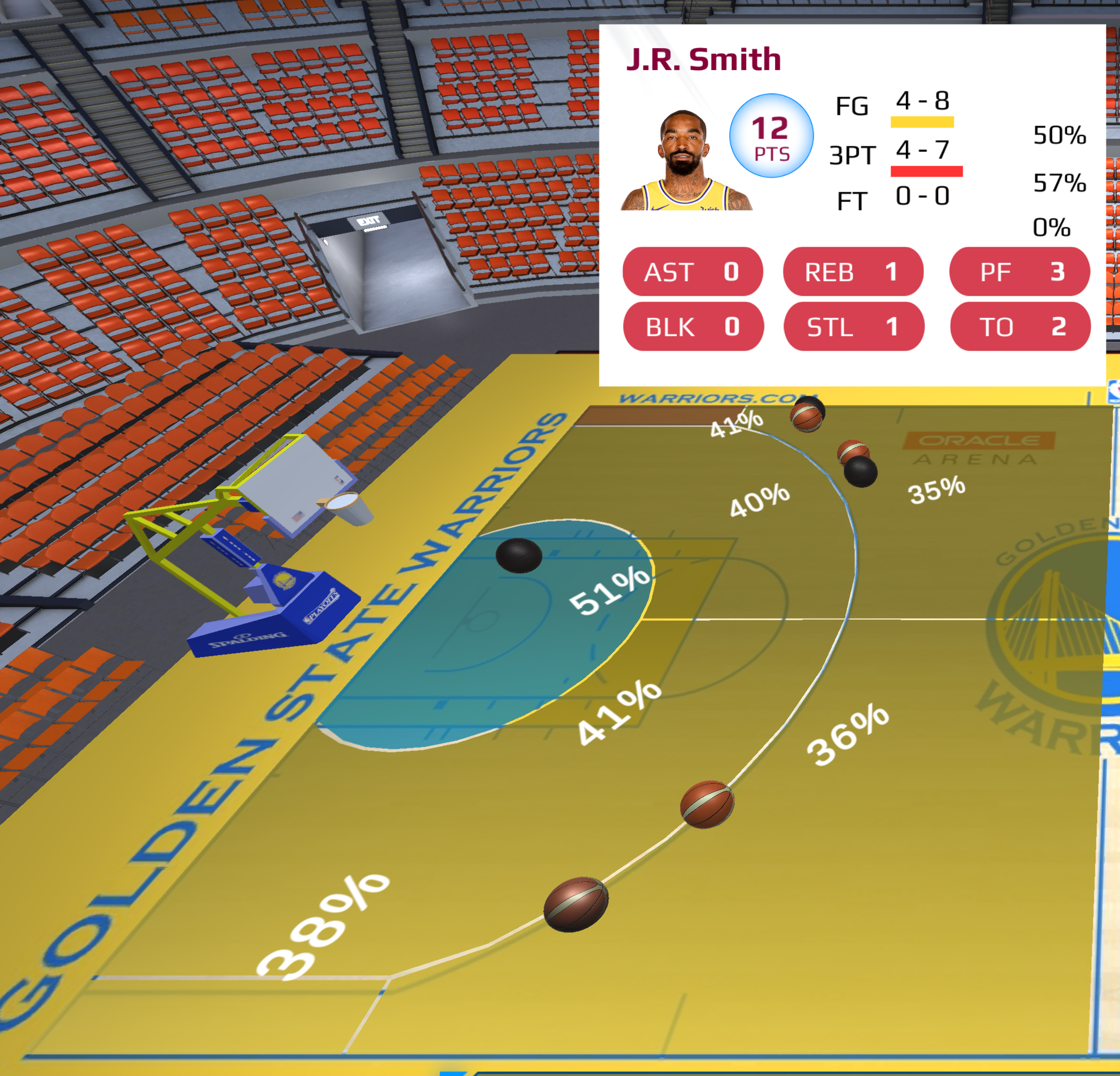}
	 \vspace{-6mm}
	\caption{Shot Chart}
	 \vspace{-0mm}
	\label{fig:ev4}
\end{wrapfigure}
We designed \ShotChart{} to have two components, 
a heat map and a player panel (\autoref{fig:ev4}). 
The heat map shows the player's seasonal shot percentage by zones (detailed in Sec.~\ref{ssec:otherdesigns}).
The areas are color-coded by the comparison to the league shot percentage average, 
from dark blue ($\leq -10\%$), blue ($\leq -5\%$), yellow ($\pm 5\%$), orange ($\geq 5\%$), and red ($\geq 10\%$). 
We also visualize shot attempts and locations with 3D basketball icons on the court to represent made (colored ball) and missed shots (black ball).
The heat map design allows fans to evaluate the shooter's detailed game performance directly within the game contexts, such as observing whether the shooter made shots in their hot zone or comparing the shot number in each zone. 
The player panel is shown on the courtside to show player box score stats, 
providing a complete picture of the player's game performance besides shooting.


\subsubsection{Team Panel}

\begin{wrapfigure}{R}{0.21\textwidth}
	\centering
	\vspace{-1mm}
	\includegraphics[width=0.21\textwidth]{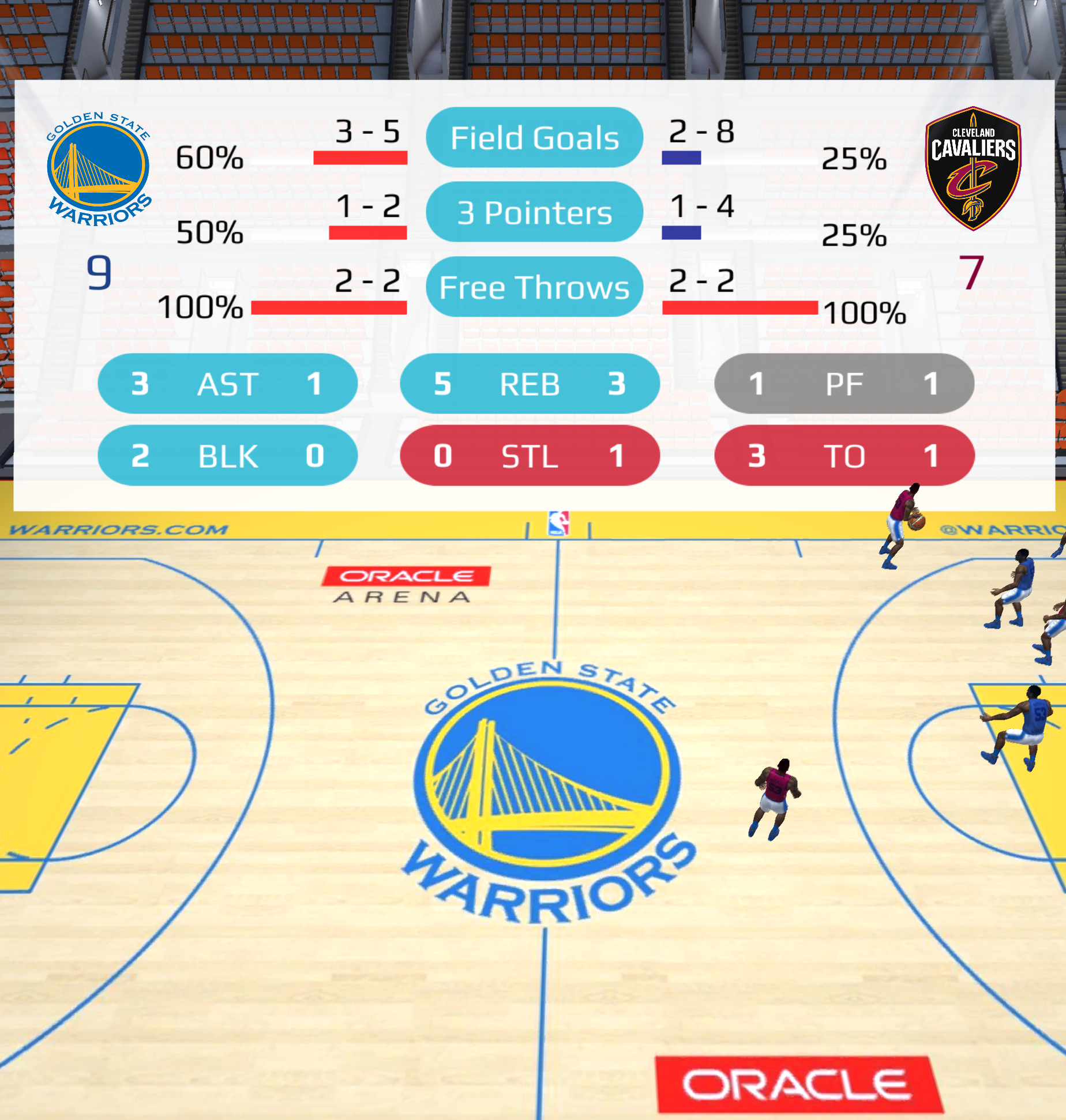}
	 \vspace{-5mm}
	\caption{Team panel}
	 \vspace{-0mm}
	\label{fig:ev5}
\end{wrapfigure}

Based on \texttt{C5 Team Comparison}, 
\TeamPanel{} is designed to provide an overview of team stats and compare team performance. 
As shown in~\autoref{fig:ev5}, 
the panel is shown at the courtside to allow instant look-ups without losing sight of the game in the front. 
Each stat label uses the color of the team with the better performance,
such as blue for 3 Pointers (Warriors have a higher shot percentage). 
The bars for each shooting stat (Field Goals, 3 Pointers, and Free Throws) are colored based on the shot percentage compared to the league's average using the same colors as Shot Chart. \re{For example,} Cavaliers have a lower 3PT shot percentage than the league's average and are coded in dark blue.
Showing a team stats comparison on a panel is common in current TV streaming services during breaks, but it usually requires cutting away from the court view.
Instead, our team panel design is always-on, without interfering with the game and thereby providing an effective direct comparison through colors.

\subsubsection{Other Design Considerations} \label{ssec:otherdesigns}

\para{Color Usage.} 
Our color encoding of shot percentages uses blue to red and is based on the convention of basketball analysis, where red indicates a hot hand (good performance), and blue indicates cold (bad performance). Similarly, Offense Trajectory uses a color scheme of reddish tones to indicate offense strengths, while Defense Form uses blue tones to show defense efforts.

\para{Shot Percentage Threshold.} 
We use the NBA's official shot charts~\cite{nbastats} and color palette to compare player and league shot percentages. 

\para{Defense Distance Threshold.} 
We define the effective defender as being within $6 \ft$ of the offense player based on the definition of an open shot~\cite{lucey2014get}.
The threshold decreases to $3 \ft$ if the offensive player has passed the defender, as a threat from the back is low. We double the threshold ($12 \ft$) for helpers as an approximation to indicate the strong side range as  there is no accurate definition for defensive helpers.

\para{Shot Chart Zones.} 
We segmented the shot chart into seven zones, including rim (within $8 \ft$), mid-range (left/right-wing), corner three (left/right), and 3-point range (left/right). We aggregated the 19 zones in the NBA shot chart~\cite{nbashotcharts} to strike a balance betwen visual complexity and usefulness. 

\subsection{Interactions}

While the five embedded visualizations can be displayed based on events,
fans still desire control over when and what data should be displayed.
During the user study, 
we applied the Wizard-of-Oz (WoZ) method~\cite{dahlback1993wizard,dow2005wizard} to allow fans to control the visualizations with voice commands, while the experimenter manually turned each visualization on and off.
The WoZ method is commonly used to evaluate complex, intelligent systems while reducing implementation effort~\cite{budzianowski2018multiwoz,palmeiro2018interaction,browne2019wizard}. 
Most importantly, the current speech recognition systems cannot meet the speed and accuracy requirements for fast-paced use cases, like sports.
We envision with the advance in speech recognition and natural language processing, a robust system will be available in the near future.
By using a WoZ method, we can ensure the consistency of user experiences without being hindered by current speech recognition systems. 
Nevertheless, 
while interaction techniques are not the focus of our study, 
we identified insights regarding users' interaction preferences as implications 
for future work in Sec.~\ref{sec:implications}.



\section{User Study}
\label{sec:user-study}

We conducted a controlled user study with active basketball fans to evaluate the usefulness and user engagement of Omnioculars.

\subsection{Experiment Set-Up \& Participants}

To simulate the game contexts,
we chose a famous game between Golden State Warriors and Cleveland Cavaliers on Dec 25th, 2015. 
This Christmas game featured the best teams from the previous season's final, 
who faced each other in the championship again, 
and was selected as one of the best games by the NBA in the 2015-16 season.
All the required data, like event data, shot locations, and player zoned shot percentage,
is publicly available~\cite{kagglenba, espn2015, nbashotcharts, basketballref},
except for the tracking data. 
We could only obtain 2D tracking data of the ball and players~\cite{sportvu}.
To preserve the authenticity of the game viewing experience in our prototype, 
we manually annotated the vertical positions and polished the player model animations to reflect the game events as closely as possible.

For our study, we selected three video clips from the simulated game. 
Each clip consists of about 30 seconds with two to three plays, focusing on shooting, offense and defense performance, and the clutch time situation. 
We used two video clips in the first part of the user study and counter-balanced them for use as either training or study clip among participants. 
We used the third clip, which contains the clutch time situation, for the second part of personalized game analysis. 

We recruited 16 participants through the university mailing list after an initial screening of their fandom levels (P1-16; M = 13, F = 3; Age: 18 - 45). We targeted active fans who reported watching games at least once a week during the game season and were interested in looking up data during the game. 
Four identified as casual fans, seven as engaged fans, and five as die-hard fans.
We conducted the user study over Zoom, which took one hour per participant to complete. 
We compensated each participant with a \$20 gift card.

\subsection{Design \& Procedures}

The study had two parts. The first part focused on assessing the understandability and usefulness of the five embedded visualizations (\ShotLabel{}, \Offense{}, \Defense{}, \ShotChart{}, and \TeamPanel{}). After being introduced to the study and filling out a consent form, the participants went through five rounds, one for each visualization condition. At the beginning of each round, we introduced each visualization design and led participants through a training task to think aloud their game insights from analyzing the embedded visualization in a game clip. After they were familiar with the visualization technique, they completed one study task of analyzing the game with the embedded visualization on another game clip. The participants were asked to rate and comment on the useful contexts supported by each visualization. 
The second part focused on the usefulness and engagement of the Omnioculars system as a whole. We asked participants to freely explore different visualization combinations using voice commands. Participants first completed a training task to analyze a game clip with interactive visualizations using voice commands. They then completed one study task of thinking aloud their game insights from using interactive visualizations in another game clip. We asked them to rate the usefulness and engagement of the overall Omnioculars system and to comment on their strategy of using different visualizations to generate game insights. At the end of the study, participants filled out a post-study questionnaire. 

\subsection{Measures}
\textbf{Part 1.} For each embedded visualization condition, we collected the subjective ratings on a 7-point Likert scale \re{from low to high}, including understandability, usefulness, engagement, and novelty. We also collected qualitative feedback, including the game insights generated by the participants while watching the game clip with each visualization, and the oral feedback about the contexts in which they found each visualization helpful and interesting to use.

\para{Part 2.} We collected subjective ratings on the usefulness and engagement of Omnioculars, where we derived our questions from prior work~\cite{o2010development}, including \textit{``It was helpful''}, \textit{``It was fun''}, \textit{``I felt in control''}, \textit{``I felt encouraged''}, and \textit{``I am likely''} to use Omnioculars to consume the desired game data in-game. We collected qualitative feedback of participants' strategy of using different visualizations and the strengths, weaknesses, and suggestions for Omnioculars in a post-study questionnaire. 

\para{Qualitative Analysis.}
Participants' responses were analyzed by two authors using affinity mapping. In Part 1, we derived primary use cases for each visualization by grouping similar user insights. In Part 2, we categorized user strategy in personalizing the visualizations and the subsequent insights. The grouping results are shown in Appendix D. 


\begin{figure}[t!]
    \centering
    \includegraphics[width=\linewidth]{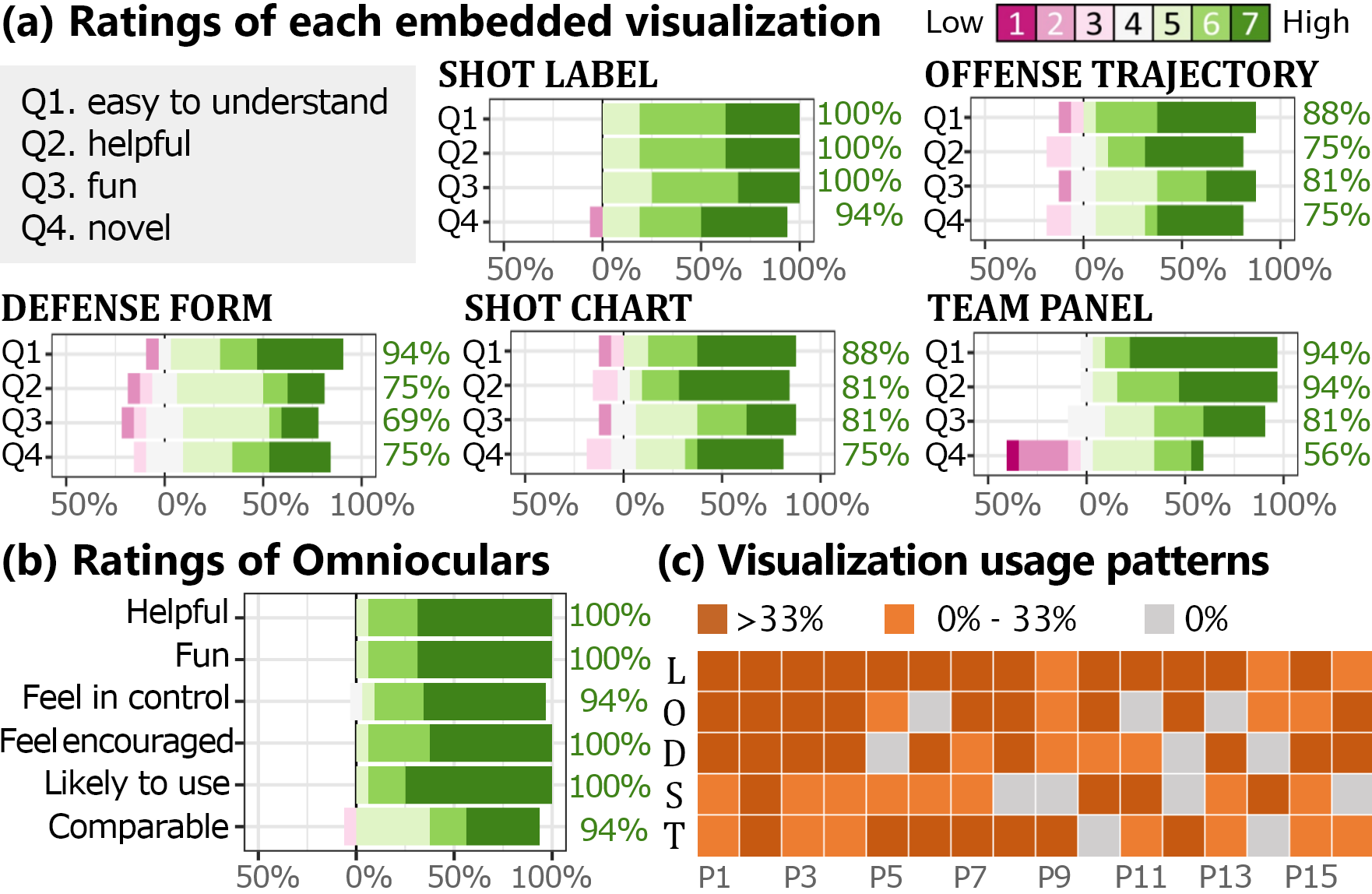}
    \vspace{-5mm}
    \caption{User study results. In Part 1, (a) participants rated all five embedded visualizations to be easy to understand, helpful, fun to use, and novel ways to present game data (Mdn $\geq 5$). In Part 2, (b) participants confirmed the usefulness and engagement of Omnioculars, were likely to use it in future games (Mdn$=7$), and perceived the simulated game as comparable to an actual game (Mdn$=6$). (c) Participants used different visualizations to perform game analysis.
    } 
    \label{fig:rating}
    \vspace{-4mm}
\end{figure}

\section{Results \& Discussion}
\label{sec:results}
We present the user study results on how well embedded visualizations help users enhance game understanding and engagement,
and discuss the design implications for future interactive game viewing.

\subsection{What Game Insights Do Fans Generate with Each Embedded Visualization?}
In Part 1 of our study, participants provided their game insights while using each embedded visualization, and commented on helpful use cases. We discuss user subjective ratings and findings below. A detailed categorization of user comments can be found in Appendix D. 

%
\para{Five embedded visualizations were perceived to be useful and novel}. 
In Fig.~\ref{fig:rating}a, participants rated each visualization from 1 (low) to 7 (high) on Q1) easy to understand, Q2) helpful, Q3) fun to use, and Q4) novel ways to present game data. The majority rated all five embedded visualizations positively ($>4$) in usefulness and engagement, \re{reporting that they were} easy to understand ($\geq $88\%), helpful ($\geq $75\%), and fun ($\geq $69\%). For novelty, Shot Label was rated novel by 94\% of the participants, Offense, Defense, and Shot Chart by  75\%, and Team Panel by 56\%.

\para{Participants used the \ShotLabel{} to evaluate individual player performance and player decisions}. Participants used color on the dynamic label to evaluate players' shot performance, \textit{``Sean Livingston is very low from the 3-point line (P2)''}. 
Furthermore, some participants contemplated the player's decision-making, \textit{``When I see Lebron drives, the color changed drastically, so I get that why he did not make the three-point shots and tried to take the drive. (P1)''}. Overall, participants found the Shot Label most helpful to focus on player and ball movement, and after shooting.

\para{Participants used the \Offense{} to spot the shot opportunity and team strategy}. Participants used the circle underneath the players to evaluate the shot opportunity, 
\textit{``They passed to people who had more space, or a larger circle, at the time (P7)''}, and the trails to read the team strategy, 
\textit{``Like this run, [trails] help me understand the team better, not just the player individually (P1)''}. 
Participants found the Offense Trajectory most useful when they focused on team offense strategy  
and following offense during the transition, \textit{``
When defense changes quickly, 
seeing opportunity arose when somebody was breaking away to the basket or slipped away from the defenders is helpful (P3)''}.

\para{Participants used the \Defense{} to track defense changes and strategy}. Participants used the color of the circle underneath the players to identify defenders. 
They also observed the defense strategy,
\textit{``The shaded area between the defenders shows how the defense blocked the paint area to get the rebound (P6)''}. Participants found the Defense Form most useful when they wanted to track defense changes during a switch or transition 
and to identify the defense strategy. 

\para{Participants used the \ShotChart{} to examine the overall shot performance and shot decisions}. Participants used the heat map and virtual balls on the court to evaluate the player's shot performance. 
They also derived insights on players' shot decisions 
and on how players compare between each other, \textit{``It’s interesting to see the chart differences between shooters and role players. (P13)''}. Participants found the Shot Chart most useful to understand shot selection, \textit{``When there were lots of shots from a particular area, to understand if they were high percentages shot or not (P11)''}, and to evaluate players' shot performance immediately, 
\textit{``to understand someone’s performance and contribution, especially with players I don’t know (P14)''}.

\para{Participants used the \TeamPanel{} to compare team performance}. Participants used the team panel to check on game stats. 
They also derived game insights from the color encoding, \textit{``Interesting to see both teams had lower shot percentages compared to the league average (P10)''}. 
Participants found the Team Panel most useful to follow the team's progress throughout the game 
and to analyze areas of focus, \textit{``When the game score is close, I can see the key factors to evaluate and predict the outcome (P8)''}.

\para{Overall, each embedded visualization provides helpful data on different game aspects for the participants}. The obtained insights and identified use cases aligned with our design contexts. 
While individual needs and knowledge levels vary, all participants were able to generate different game insights from each visualization.

\subsection{How Do Fans Personalize Omnioculars to Enhance Game Understanding and Engagement?}
In Part 2, participants interacted with Omnioculars to use different visualizations to generate game insights. We were interested to see how they combined or altered visualizations during the game, how their individual preferences impacted their strategies, and how the selected embedded visualizations helped users generate game insights. 

\para{Participants used various combinations of the different visualizations.}
Fig.~\ref{fig:rating}c shows the usage patterns of the five visualizations for each participant (the amount of time a certain visualization was displayed).
On the individual level, each participant chose different visualization combinations and focused on different aspects of the game. Six participants used all five visualizations, and all users used at least three. Each visualization was used frequently ($>33\%$) by several users each, i.e., Shot Label (by 13 users), Offense (9), Defense (9), Shot Chart (4), and Team Panel (7).

\begin{figure}[t!]
  \centering
  \includegraphics[width=\linewidth]{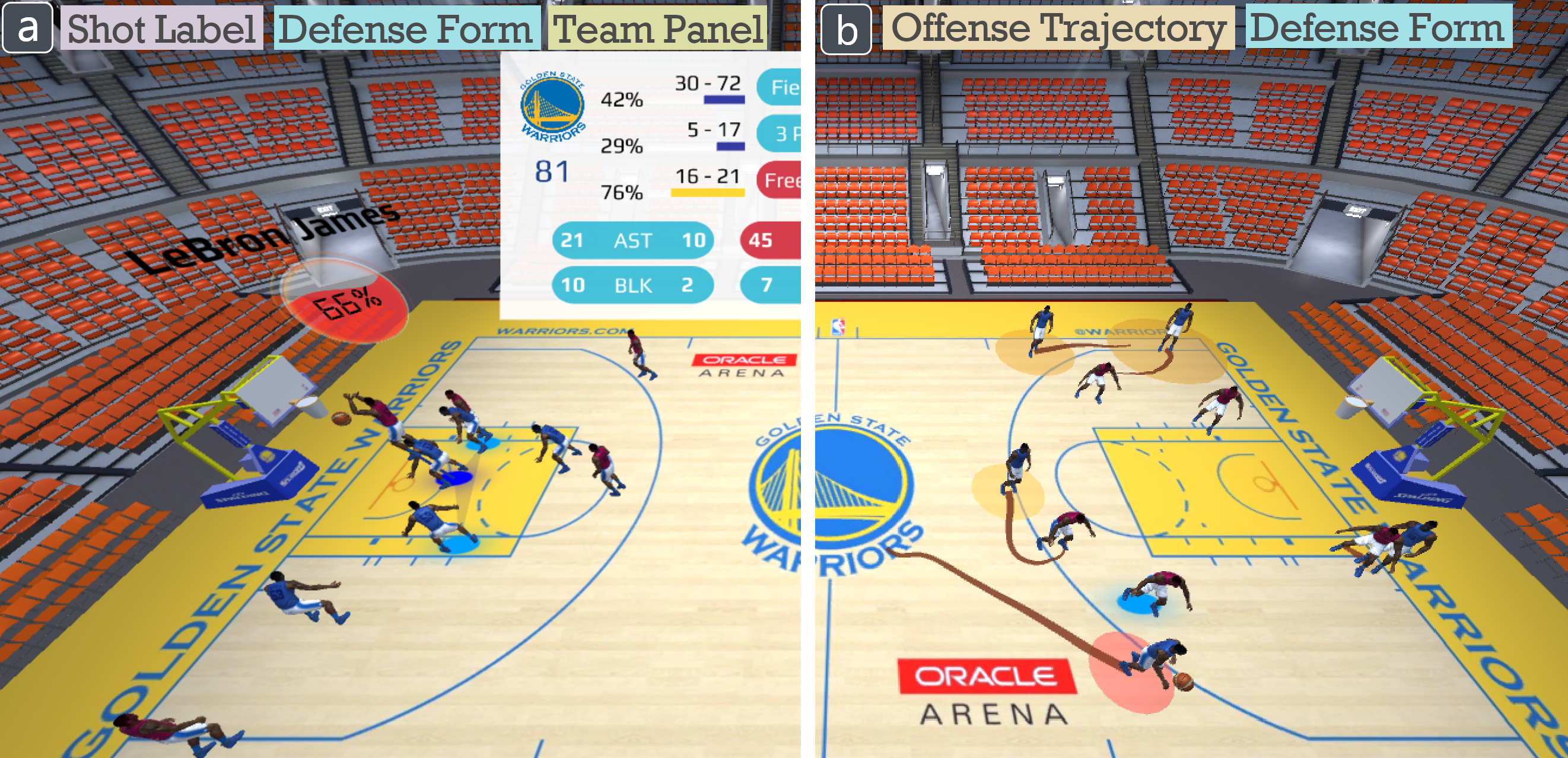}
  \vspace{-6.5mm}
  \caption{(a) and (b) show personalized visualizations configured by different users during the user study.}
  \vspace{-6.5mm}
  \label{fig:strategy}
\end{figure}

\para{Participants developed different strategies to engage with the visualizations based on individual preferences and their game focus.}
We observed two primary user strategies. 
The first strategy was \textbf{``configure a fixed set''}, \re{which was used} by eight participants. These users first identified their preferred combination of visualizations and used this fixed set throughout the game clip. They hoped to access personalized visualizations to facilitate game analysis without having to actively control them. Among them, we identified two styles of rationales, analytic-driven and experience-driven. Analytic-driven users decided the visualization set based on their game analysis needs.
For example, P6 used Team Panel, Defense, Shot Label all together (Fig.~\ref{fig:strategy}a), and explained that \textit{``these help me focus on why they are playing this way, see if they switch or not, and track the shot percentage''}. On the other hand, P3 used Team Panel at first, 
and used Shot Label and Shot Chart together for the rest of the game clip. P3 explained that \textit{``I wanted the team panel first to get an idea of how things were going thus far.
Shot Label and Shot Chart are very useful for seeing current instant threats by the ball handler. I also appreciated seeing whether a player was hot compared to their average''}. 
Experience-driven users chose the visualizations based on engagement. For example, P2 wanted to start with everything on because \textit{``it's already such a fast-paced sport that I liked just having them on.''}
P15 used Shot Label, Defense, and Shot Chart together because \textit{``they are very fun to use and watch''} (Fig.~\ref{fig:teaser}).
The second strategy was \textbf{``alter based on context''} \re{adopted} by eight users. 
Participants selected visualizations based on their needs, which varied based on the current game context.
For example, P5 explained that \textit{``during the clutch time, I'll look at team panel. When the game is not as intense, I will look at defense and offense''}. Some participants selected visualizations based on the team they support, \textit{``If I'm watching a game of a team I am rooting for, I focus on the visualization on that (offense/defense vis when the team is attacking/defending)''}. Other fans showed more flexibility, \textit{``I first use Offense and Defense to understand the team drills} (Fig.~\ref{fig:strategy}b). \textit{When I am familiar with them, I'll use Shot Label to evaluate if the shot selection is reasonable (P9)''}. 

\para{Embedded visualizations help participants derive insights.} 
When asked about the game insights they obtained with Omnioculars, participants focused on distinct aspects, such as \textit{identify} player movement and decision making, \textit{compare} player's in-game stats, or \textit{summarize} game performance. These goals align with the data and tasks we identified in Sec.~\ref{ssec:data_context}.
For example, P2 used the Offense Trajectory to focus on player decisions, \textit{``when he just passed, I think it's cool to see that there's not really open shots as there's no big yellow circle anywhere''}. P3 evaluated player's shooting with the Shot Label, stating \textit{``it's so interesting to see the shot percentage climb up as they drive to the basket''}.
Additionally, 
P1 used the Team Panel to derive a game summary, \textit{``We are doing really well in the assist and block, wow that's unexpected. The turnover is a problem, but if you have more assists, you're supposed to have more turnovers. I think that's acceptable.''}.

%




\subsection{Implications for Future Interactive Game Viewing}
\label{sec:implications}


\para{Flexible embedded visualization design is important to enable a customized experience.}
A key finding is that individual needs for data and desired interaction levels vary. Based on the user's preferences and basketball experience, some people felt a visualization was distracting and overwhelming, while others felt the same visualization helped them see the data more easily. For example, the trails in the Offense Trajectory visualization were perceived useful to evaluate offense strategy by some die-hard fans, \textit{``The trails are great to see the players' strategy when the players are closer, like pick-and-roll. It makes it easier to see the tactics (P9)''}. 
However, other fans felt they only focused on the circles to identify the shot opportunity and did not utilize the trails, \textit{``I just feel like it's too much information to follow. (P2)''}. 
A recommended solution is to decouple the individual components of a visualization, such as separating the trail and circle in the Offense Trajectory visualization, or the heat map and player panel in the Shot Chart visualization. As participants focus on a range of goals based on their preferences, allowing them to select and mix from independent visual components can best fulfill their individual needs. 

\vspace{-0.8mm}
\para{Embedded sports visualizations are still underexplored.}
We iterated our designs with targeted users, and we believe they are the most suitable in the scope of this project.
However, we also believe there is a need for more refined embedded visualizations targeting specific tasks and contexts based on our proposed design framework.
Additionally, designing embedded sports visualizations not only requires consideration for objective dimensions such as perceptual effectiveness, but also for subjective dimensions, such as engagement, which plays a more vital role compared to traditional visualization design.

\vspace{-0.5mm}
\para{Expressive interactions allow users to go beyond a passive viewing experience.}
We used voice commands as the interaction mechanism for our study, as we believe, with overwhelming choices, it can best articulate the user's intention~\cite{srinivasan2020ask,srinivasan2021collecting,Shen2022TowardsNL}, especially in fast-paced sports watching. 
Participants had different opinions about the provided interactions:
a few participants preferred to have a certain level of default behaviors, \textit{``I wish Shot Chart would turn on automatically for a few seconds after shots 
(P2)''} and \textit{``I feel like it would be a hassle to be turning it on and off and on and off (P3)''}, while others enjoyed having interaction, 
\textit{``I liked the voice activated part of it, very useful to follow the game and still visualize what I wanted (P11)''}. To this end, we suggest an interaction approach that would allow users to configure default settings based on their game focus on desired data. 
Future work can explore interaction design space in the spectrum from manual to automatic and train personalized AI models to support semi-automatic interactions. 
In terms of functionality, we only allowed simple interactions to toggle visualizations on and off, mainly considering the fast-paced nature of sports. However, more complicated interactions could be required (check the average stats of three players). 
Future research can explore other interactions to support in-game data analysis.

\para{A systematic evaluation of human perception and attention could shape embedded visualization design.}
Participants highlighted the advantages of our embedded visualization design for game viewing, including providing immediate feedback, being easily accessible while watching the game video, and alleviating mental load.
For example, the Shot Label visualization provides \textit{``immediate visual feedback after the pass. I've seen the color flip from red to blue to immediately know that (P3)''}. With Team Panel sitting next to the court, \textit{``I can look at it, for as long as I need to and then switch over to looking at what's happening now (P2)''}.
The visualization can help fans focus on the ball, \textit{``I could still be watching the ball, but out of the corner of my eye could see whether or not they had options (P12)''}. 
As the first step towards understanding the use of embedded visualization in sports, our work aims to provide proof-of-concept results for the design space and its effectiveness. 
As a result, we did not systematically evaluate these potential advantages in our study.
We believe, as future work, obtaining empirical knowledge of these effects can largely facilitate the development of embedded sports visualization. 

\para{\re{Future broadcasting with interactive embedded visualizations.}}
\re{
In current live streams of games, camera angles may change and cut off the scene. Without access to camera configurations and the complete video footage, embedding visualizations into videos accurately can be difficult. 
However, broadcasters already collect all the required information, such as tracking data and the full video footage. Therefore, we believe our design framework and visualizations can be adopted and extended to augment actual live games by the broadcasters, or by researchers once the data are made available. Similar to our envisioning direction, a few personalizations are already offered by the NBA, such as customized viewing angles with the NBA League Pass} \cite{nbaleaguepass} 
\re{and augmented visual overlays on the live game view by CourtVision} \cite{clipperscourtvision}.

\para{Beyond-the-screen game viewing.}
Many people prefer watching games in person to experience the live atmosphere.
Unlike watching a game through digital devices with visual effects, it is challenging to overlay extra information on a physical court.
Fortunately, with recent advances in augmented reality (AR), it is now possible to embed digital graphics into the physical world.
We believe that using AR to embed visualizations into an in-person game will enhance the game viewing experience, and help spectators make sense of the real-time game data.
We also believe our results can be largely generalized to an AR in-person game viewing scenario, as we designed and implemented our system in a 3D environment (Unity3D).
However, using our designs in AR still differs from a flat screen and we need to consider: the effect of stereoscopy vision and depth perception; the limited field of view (FoV) in current AR devices; and the ability to freely change viewing direction so that physical and digital objects can be out-of-view.
In the future, we plan to adapt our current system to AR (by adding out-of-view notifications and by simplifying visualization designs based on the FoV) and to evaluate its effectiveness.









\subsection{Simulating Basketball Games in a Virtual Environment}
\label{sec:game-simulation}
%

Our study shows that simulation can be a promising solution to ease the development and evaluation 
of visualizations embedded in physical contexts. 94\% of all users considered game insights derived from the simulated game comparable to an actual game (Fig.~\ref{fig:rating}b). 
Compared to traditional visualizations presented in digital environments,
embedded and situated visualizations presented in physical contexts
are difficult to develop and evaluate 
due to the limited accessibility and controllability of the physical contexts.
Moreover, unlike web-based visualizations,
embedded and situated visualizations are inherently challenging to distribute, reproduce, and benchmark,
hindering the research progress in this emerging direction.
By compromising a certain degree of fidelity,
simulation enables a highly accessible and controllable physical context that can be shared, reproduced, and compared universally,
significantly lowering the barrier to designing embedded or situated visualizations.
In recent years, thanks to the proliferation of low-cost VR devices,
many researchers~\cite{shimizu2019sports, tsai2020feasibility, tanaka2018scope, ye2020shuttlespace, chuTIVEEVisualExploration2022, DBLP:conf/uist/ChengY0SL21}
have leveraged VR to simulate real-world environments
in user-centered design studies.
We envision and advocate a standard simulation protocol for developing visualizations in physical contexts,
which may eventually facilitate the research of embedded and situated visualizations. 

\section{Conclusion \& Future Work}
This study explored the design space of embedded visualization for basketball in-game data analysis. 
Through a user-centered design process, we presented a formative study of basketball fans' in-game data needs and proposed a context-driven design framework of four design elements --- \emph{Scenario}, \emph{Data}, \emph{Task}, and \emph{Embedded Vis}. We designed five embedded visualizations for five primary game contexts and developed Omnioculars, a prototype for personalized game viewing with embedded visualizations. 
Our evaluation results suggest that each embedded visualization provides helpful data on different game aspects, and fans developed different strategies to engage with visualizations based on their individual preferences and game focus with Omnioculars.

%
%
For future work, we identified the necessity to allow customizable visual complexity and interaction with embedded sports visualizations based on individual preferences and knowledge levels. We also hope to build upon the current design framework to explore using embedded visualizations for immersive and situated in-game data analysis.

\scriptsize
\acknowledgments{This work is
supported by NSF grants III-2107328 and IIS-1901030.
We thank Hans H., Joan C., Harvey H., Ahmed S., Nhan H., and Max M. for their time. }

\bibliographystyle{abbrv}
\bibliography{main}
\end{document}